\documentclass[preprint,aps,showpacs]{revtex4}
\begin{document}
%\draft
\preprint{UCI-TR 2007-2}
\title{Noncommutative Coordinates Invariant under Rotations and Lorentz Transformations}
\author{Myron Bander\footnote{Electronic address: mbander@uci.edu}
%\addtocounter{footnote}{3}%
}
\affiliation{
Department of Physics and Astronomy, University of California, Irvine,
California 92697-4575}

\begin{abstract}
Dynamics with noncommutative coordinates invariant under three dimensional rotations or, if time is included, under Lorentz transformations is developed. These coordinates turn out to be the boost operators in $SO(1,3)$ or in $SO(2,3)$ respectively. The noncommutativity is governed by a mass parameter $M$. The principal results are: (i) a modification of the Heisenberg algebra for distances smaller than $1/M$, (ii) a lower limit, $1/M$, on the localizability of wave packets, (iii) discrete eigenvalues of coordinate operator in timelike directions, and (iv) an upper limit, $M$, on the mass for which free field equations have solutions. Possible restrictions on small black holes is discussed.
\end{abstract}

\maketitle
\section{Introduction}\label{intro}
Recently there has been significant interest in extending quantum mechanics and quantum field theory from ordinary coordinates to noncommutative geometries. Much of the work has implemented such noncommutativity, $[x_a,x_b]=i\theta_{ab}$, through the Groenewold-Moyal \cite{GroenMoy} star product wherein the ordinary product of two functions is replaced by
\begin{equation}\label{starprod}
f(x)\star g(x)=\left .\exp\left(i\frac{\theta^{ab}}{2}
  \partial^{(x)}_a\partial^{(y)}_b
  \right )f(x)g(y)\right |_{y=x}\, .
\end{equation}
As specific directions are singled out, this procedure is not Lorentz invariant and in greater than two dimensions not even rotationally invariant. This product does respect a ``twisted" Poincar\'{e} invariance \cite{Chaichian:2004za, Wess:2003da}; implications of such invariance for field theories have been discussed \cite{Fiore:2007vg} in great detail recently. 

A different approach, in which the position coordinates are replaced by operators that have non trivial commutation relations among each other and under rotations transform into each other preserving these commutation relations has been pursued by this author \cite{Bander:2005bg}. In that paper the space coordinates are represented by operators acting on coherent states based (in three space dimensions) on the group $SO(1,3)$.  Although in this work we shall again be interested in this group, the approach now is different. Specifically, we will formulate a dynamics with noncommutative coordinates that transform as vectors under rotations or, with the introduction of time, under the Lorentz group; this dynamics, at least at the level of free fields, is invariant under modified translations and a full Poincar\'e invariance can be implemented.  The caveat of restricting the claim of Poincar\'e invariance to free field theories is that usual formulations of interacting ones involve time ordered products.  Invariance of such products can be implemented only if commutators of operators vanish for space-like separations \cite{Alvarez-Gaume:2001ka,Doplicher:2006vy, Balachandran:2006ib}; as in the present formulation time will be associated with an operator, the notion of time ordering is unclear.

The motivation for the procedure used in the present work is as follows. If we assume that spatial coordinates have commutation relation of the form
\begin{equation}\label{simplecomm}
[X_i,X_j]=iF_{i,j}\, ,
\end{equation}
where $F_{ij}$ is  non constant and antisymmetric, rotation invariance demands that it transforms as a tensor; if we demand that (\ref{simplecomm}) is to be time independent then a possible choice is $F_{ij}\sim {\cal M}_{ij}$, where ${\cal M}_{ij}$ is the angular momentum tensor. Requiring further that the $X_i$'s transform appropriately under time reversal and under parity leads us to postulate that $X_i\sim {\cal K}_i$, where the ${\cal K}$'s are the boost operators of the group $SO(1,3)$; even though the group $SO(1,3)$ appears, this formulation is invariant {\em only} under rotations; a true extension to relativistic invariance will be given in Sect.~\ref{Minkowski}. Details of this generalization of position operators to ones with with non trivial commutation relations are presented in Sect.~\ref{Euclidean}. The scale of noncommutativity is governed by a mass parameter $M$ and in the limit $M\rightarrow\infty$ ordinary three dimensional quantum mechanics is recovered. The Hilbert space these operators act on is a momentum space which serves as a basis for the simplest representation of the Poincar\'{e} group, namely the one for spinless particles of mass $M$. For $M$ finite and for momenta larger than $M$ and/or distances smaller than $1/M$ the Heisenberg commutation relations between position and momentum are modified; this, in turn, forces a modification of the concept of space translations. A procedure for obtaining functions of these noncommuting coordinates and their derivatives are also discussed in this section. Functions are related to their commuting counterparts by having common Fourier transforms and commutators of these functions with momentum operators serve as derivatives. The definition of an  integral of a product of such functions, consistent with the previously discussed procedure for taking derivatives, is subtle but can be implemented as a specific matrix element of such a product. An interesting consequence of such coordinate noncommutativity is that fluctuations in the measurement of $\sqrt{X^2}$ must exceed $1/M$. Relative and center of mass coordinates needed for two body problems are presented.

Noncommuting coordinates invariant under $SO(3)$ were identified with boosts of $SO(1,3)$. In Sect.~\ref{Minkowski} this procedure is extended to noncommuting space and time coordinates that transform under a the $SO(1,3)$ Lorentz group; this leads us to identify these noncommuting space-time coordinates with boosts of the $SO(2,3)$ anti-de-Sitter group. Such an identification was suggested by Snyder \cite{Snyder:1946qz} in the first discussion of this problem. The operators for spacial coordinates have continuous eigenvalues but those of $\hat q\cdot X$, where $\hat q$ is timelike, are {\em discrete} with separations of $1/M$. As in the previous section functions, derivatives and integrations are discussed. Definition of integration permits us to introduce in Sect.~\ref{FieldTheory} a Lagrangian and  an action for a free scalar field theory, where the field is a functions of the noncommuting space-time. A significant consequence is that solutions of the equations of motion for such a free field theory exist only for masses $m\le M$.

The above limit on masses of particles combined with the bound on the minimum size of spatial wave packets may put restrictions on the existence of small black holes. This and other results are summarized in Sect.~\ref{Conclusions}. 

\section{Euclidean Noncommuting Coordinates} \label{Euclidean}
The discussion in this section will focus on noncommuting coordinates in three spatial dimensions invariant under $SO(3)$; these coordinates will be identified with boosts in an $SO(1,3)$ group. An extension to $d$-dimensional noncommuting coordinates invariant under $SO(d)$  and embedded in $SO(1,d)$ is straightforward. 
\subsection{Coordinates}
The usual, commuting, position variables, $x_i$, acting on momentum states $|{\vec p}\rangle$, with $\langle {\vec p}\, '|{\vec p}\rangle=\delta({\vec p}\, '-{\vec p})$, have the form
\begin{equation}\label{ordpos}
x_i=i\frac{\partial}{\partial p_i}\, .
\end{equation}
As previously mentioned, the noncommuting position operators will be related to boost operators of an $SO(1,3)$ Lorentz group. The generators of the full Poincar\'{e} group,  and in turn the Lorentz group, can be represented as operators acting on the momentum states; we are interested in an irreducible representation of the Poincar\'e group \cite{Wigner} with a mass $M$ and spin zero. To this end we define
\begin{equation}\label{p0def}
p_0=\sqrt{{\vec p}\cdot{\vec p}+M^2}\, ,
\end{equation}
which we use to construct the boost operators
\begin{equation}\label{boost}
{\cal K}_i=\sqrt{p_0}\left(i\frac{\partial}{\partial p_i}\right)\sqrt{p_0}\, .
\end{equation}
It is easy to check that these satisfy the requisite commutation relations
\begin{equation}\label{boostcomm}
\left[{\cal K}_i,{\cal K}_j\right]=-i{\cal M}_{ij}\, ,
\end{equation}
with ${\cal M}_{ij}$ the angular momentum,
\begin{equation}\label{angmom}
{\cal M}_{ij}=i\left(p_j\frac{\partial}{\partial p_i}-p_i\frac{\partial}{\partial p_j}\right)\, ;
\end{equation}
in three dimensions ${\cal M}_{ij}=\epsilon_{ijk}J_k$. (Although we shall not use these, similar realizations of the other representations of the Poincar\'{e} group for positive $M^2$, the ones with internal spin, exist. If total angular momentum is given by the right hand side of (\ref{angmom}) plus the spin part $S_{ij}$ the boost is obtained by adding $-S_{in}p_n/(p_0+M)$ to the right hand side of (\ref{boost}).)

The {\it noncommuting} coordinates we shall use are
\begin{equation}\label{noncommpos}
X_i=\frac{{\cal K}_i}{M}=\frac{1}{M}\left({\vec p}\cdot{\vec p}+M^2\right)^{\frac{1}{4}}\left(i\frac{\partial}
      {\partial p_i} \right)\left({\vec p}\cdot{\vec p}+M^2\right)^{\frac{1}{4}}\, .
\end{equation}
Their commutation relations follow from (\ref{boostcomm}),
\begin{equation}\label{poscomm}
\left[X_i,X_j\right]=-\frac{i}{M^2}{\cal M}_{ij}\, .
\end{equation}
The mass $M$ plays the role of the noncommutativity parameter. From (\ref{noncommpos}) we see that in the limit  $M\rightarrow\infty$, $X_i\rightarrow x_i$ and for distances greater than $1/M$ ordinary commuting geometry is recovered. In subsequent parts of this work we shall denote commuting coordinates by lower case letters and noncommuting ones  by corresponding upper case letters; to emphasize that in the noncommuting situation momentum operators are as they were in the commuting one lower case letters will continue to be used for these.

The commutation relations of the boost operators with momenta
\begin{eqnarray}\label{K-pcomm}
\left[{\cal K}_i,p_j\right]&=&i\delta_{ij}p_0\, ,\nonumber\\
\left[{\cal K}_i,p_0\right]&=&ip_i
\end{eqnarray}
yield a modified position-momentum commutator
\begin{equation}\label{pXcomm}
\left[p_i,X_j\right]=-i\delta_{ij}\frac{p_0}{M}\, ;
\end{equation}
again, in the limit $M\rightarrow\infty$ we recover the standard Heisenberg algebra. The extra factor of $p_0/M$ in (\ref{pXcomm}) will affect the implementation of coordinate translations. 

It is interesting to note that expanding (\ref{pXcomm}) to first order in $(p/M)^2$ yields corrections to the Heisenberg commutation relation that have been postulated, as has (\ref{poscomm}) to the same order, \cite{Kempf:1996fz} in order to explain a minimum length that appears in string theory \cite{Kempf:1996nk}. (In the notation of ref. \cite{Kempf:1996fz}, $M^2=1/2\beta$ and $\beta'=0$). Some consequences of these modifications of the algebra of quantum mechanics are presented in \cite{Kempf:1996nk}.

\subsection{Functions, Differentiation and Integration}\label{funct1}
The relation between functions $f(\vec x)$ of commuting coordinates $x_i$ and corresponding functions of the noncommuting ones is achieved through their Fourier transforms,
\begin{equation}\label{f(x)tof(X)}
f(\vec x)=\int d{\vec q}\tilde{f}(\vec q\,)e^{i\vec q\cdot \vec x}\rightarrow 
  f(\vec X)=\int d\mu_3({\vec q})\tilde{f}(\vec q\, )e^{i\vec q\cdot \vec X}\, .
\end{equation}
$\mu(\vec q)$ is the group invariant measure 
\begin{equation}\label{invmeas1}
d\mu_3(\vec q)=\left(\frac{\sinh(q/M)}{q/M}\right)^2d{\vec q}\, .
\end{equation}
As $i\left[p_i,f(\vec x)\right]=\partial_if(\vec x)$, we may carry this over to a definition of derivatives of $f(\vec X)$,
\begin{equation}\label{diffdef}
\partial_if(\vec X)=i\left[p_i,f(\vec X)\right]\, ;
\end{equation}
this prescription satisfies the Leibniz rule for the differentiation of a product of functions of the noncommuting coordinates. 

The integral over all space of a product of functions of ordinary coordinates is a convolution of their Fourier transforms,
\begin{equation}\label{ordint}
\int d{\vec x} f_1(\vec x)\cdots f_n(\vec x)=(2\pi)^3\int d{\vec q}_1\cdots d{\vec q}_n
  {\tilde f}_1({\vec q}_1)\cdots {\tilde f}_n({\vec q}_n)\delta({\vec q}_1+\cdots +
    {\vec q}_n)\, .
\end{equation}
As $\exp(i{\vec q}\cdot{\vec x})|\vec k\rangle=|{\vec k} -{\vec q}\rangle$, an expression for the $\delta$-function is
\begin{equation}\label{deltadef}   
\delta({\vec q}_1+\cdots +{\vec q}_n)=\langle \vec k|e^{i{\vec q}_1\cdot{\vec x}}\cdots
     e^{i{\vec q}_n\cdot{\vec x}}|\vec k\rangle\, ,
\end{equation}
valid for any state $|\vec k\rangle$.   We shall use the above as a guide for the definition of integration of functions of the noncommuting coordinates.
\begin{equation}\label{noncommint}
``\int d{\vec X}" f_1(\vec X)\cdots f_n(\vec X)=(2\pi)^3\int d\mu_3({\vec q}_1)\cdots d\mu_3({\vec q}_n)
  {\tilde f}_1({\vec q}_1)\cdots {\tilde f}_n({\vec q}_n)\langle \vec k|e^{i{\vec q}_1
  \cdot{\vec X}}\cdots e^{i{\vec q}_n\cdot{\vec X}}|\vec k\rangle\, .
\end{equation}
As any state $|\vec k\rangle$ is obtainable from any other by a unitary transformation and noting that the measure for the $q$ integrations is invariant under such transformations, the above definition for different $\vec k$'s are equivalent; using $|\vec k=0\rangle$  simplifies calculations and we shall use this state from now on and drop the quotation marks in (\ref{noncommint}).  Derivatives defined by (\ref{diffdef}) insure the desired relation $\int d{\vec X} \partial_if(\vec X)=0$. Two examples are in order.
\begin{equation}\label{eiqX}
\langle\vec 0|e^{i\vec q\cdot\vec X}|\vec 0\rangle=\delta(\vec q)\, ,
\end{equation}
as for ordinary functions. For the two point function the result is
\begin{equation}\label{eiqq'X} 
\langle\vec 0|e^{i\vec q\cdot\vec X}e^{-i\vec q\, '\cdot\vec X}|\vec 0\rangle=\delta(\vec q-\vec q\, ')\left[\frac{q/M}{\sinh \left(q/M\right)}\right]^2\, ,
\end{equation}
with the term multiplying the $\delta$ function canceling a similar one in the invariant measure (\ref{invmeas1}).
For $f(\vec X)$ and $g(\vec X)$ related to $f(\vec x)$ and $g(\vec x)$ by (\ref{f(x)tof(X)})  we find that  
\begin{equation}\label{intrel}
\int d\vec X f(\vec X)g(\vec X)=\int d\vec xd\vec y f(\vec x)g(\vec y) \, ;
\end{equation}
this is no longer true for integrations of a product of three or more functions. 

\subsection{Localization and Translations}\label{LocTrans}
As the coordinates $X_i$ do not commute with each other, a state with a precise position does not exist.  It is still interesting to ask what are the eigenstates of $\hat r\cdot\vec X$ for a specific unit vector $\hat r$ and what is the minimum eigenvalue of $\vec X\cdot\vec X$. The eigenstates of $\hat r\cdot\vec X$ with eigenvalues $r$ are
\begin{equation}\label{positeigen}
\psi_{\hat rr}(\vec p)=\sqrt{\frac{M}{2\pi p_0}}\left(\frac{p_0-\hat r\cdot\vec p}
      {p_0+\hat r\cdot\vec p}\right)^{\frac{iMr}{2}}\, ;
\end{equation}
the normalization is $\int d({\hat r\cdot\vec p})\psi^{\dag}_{\hat rr'}(\vec p)\psi_{\hat rr}(\vec p)
=\delta(r-r')$ and, as expected, these approach $\exp(-ir\hat r\cdot\vec p)/\sqrt{2\pi}$ as $M\rightarrow\infty$. For different directions $\hat r$ these functions are not orthogonal to each other. 
A complete set of commuting operators that includes $\hat r\cdot\vec X$ contains, in addition to this position operator, the two vector ${\vec p}_\bot$, where ${\vec p}_\bot\cdot\hat r=0$. For a given $\hat r$, the states $|{\vec p}_\bot,\hat rr\rangle$ are complete and under rotations transform to a similar set with a rotated $\hat r$. 

The eigenvalues of $X^2$ control the extent to which a packet can be localized in position. A lower bound on such eigenvalues may be obtained by noting that the $SO(1,3)$ Casimir operator ${\cal K}^2 -J^2$ 
equals $\rho^2-j_0^2+1$ \cite{Representation} for representations labeled by $(\rho, j_0)$, with real $\rho\ge 0$, and with all angular momenta in the representation having values greater than $ j_0$. As $X^2=\left({\cal K}^2-J^2+J^2\right)/M^2$, its eigenvalues are $\left[\rho^2+1-j_0^2+j(j+1)\right]/M^2$, with $j\ge j_0$; thus we find that $X^2\ge 1/M^2$ and wave packets cannot be localized to better than $1/M$.

This inability to localize reflects itself in the nonexistence of a translation operator taking $\vec X$ to $\vec X+\vec a$, with $\vec a$ a c-number.  It is easy to show that a unitary operator $U(\vec a)$ with the property $U^\dag(\vec a){\vec X}U(\vec a)={\vec X}+{\vec a}$ does not exits. If it did then (\ref{poscomm}) would imply that $\left[U(\vec a),{\vec J}\right]=0$ which cannot be as the angular momentum must rotate the translation vector $\vec a$.  

The operator $\exp(i\vec a\cdot\vec p)$, which does translate the commuting coordinate vectors $\vec x\rightarrow \vec x+\vec a$, has the following effect on the noncommutative vectors $\vec X$,
\begin{equation}\label{translations}
e^{i\vec a\cdot \vec p}\vec Xe^{-i\vec a\cdot \vec p}=\vec X+\vec a\, \frac{p_0}{M}\, .
\end{equation}
Thus, for $p<<M$ or distances larger than $1/M$ we recover the usual translations, while for distances smaller than $1/M$ the limit on localization of position wave packets makes translations fuzzy. However, as we shall see in the next section, a form of two body interactions invariant under overall translations does exist. The integration defined in (\ref{noncommint}) {\em is invariant} under translations described by (\ref{translations}).

\subsection{Two Body Interactions}
One body dynamics can be reformulated for the noncommuting coordinates. As outlined in Sect.~\ref{funct1}, this is achieved by replacing the  potential $V(\vec x)$ by $V(\vec X)$. Two body interactions present a problem in that replacing $V(\vec x^{(1)}-\vec x^{(2)})$ by the corresponding $V(\vec X^{(1)}-\vec X^{(2)})$ (superscripts refer to the two particles) does not permit a separation into mutually commuting center of mass and relative coordinates. This is a consequence of the lack of a simple translation operator, as was discussed at the end of Sect.~\ref{LocTrans}.  With the usual definitions of relative and center of mass coordinates,  $\vec p_{\rm rel}=(m^{(2)}\vec p^{\, (1)}-m^{(1)}\vec p^{\, (2)})/(m^{(1)}+m^{(2)})$, $\vec x_{\rm rel}=\vec x^{\,  (1)}-\vec x^{\, (2)}$ and $\vec p_{\rm cm}=\vec p^{\, (1)}+\vec p^{\, (2)}$, $\vec x_{\rm cm}=(m^{(1)}\vec x^{\, (1)}+m^{(2)}\vec x^{\, (2)})/(m^{(1)}+m^{(2)})$, we use the procedure outlined in (\ref{noncommpos}) for obtaining the noncommuting version of these, namely multiply the position operators on the left and right by $(\vec p\cdot\vec p+M^2)^{\frac{1}{4}}$, where $\vec p$ is respectively $\vec p_{\rm rel}$ or $\vec p_{\rm cm}$; thus instead of $\vec X^{(1)}-\vec X^{(2)}$ as a relative coordinate, we define
\begin{eqnarray}\label{cm-rel}
X_i^{\rm rel}&=&\frac{i}{M}\sqrt{\vec p_{\rm rel}\cdot\vec p_{\rm rel}\cdot+M^2}
       \left(\frac{\partial}{\partial p_i^{(1)}}-\frac{\partial}{\partial p_i^{(2)}}\right)
           \sqrt{\vec p_{\rm rel}\cdot\vec p_{\rm rel}\cdot+M^2}\, ,\nonumber\\
&{}&\\
X_i^{\rm cm}&=&\frac{i}{M(m^{(1)}+m^{(2)})}\sqrt{\vec p_{\rm cm}\cdot\vec p_{\rm cm}\cdot+M^2}
       \left(m^{(1)}\frac{\partial}{\partial p_i^{(1)}}+m^{(2)}\frac{\partial}{\partial p_i^{(2)}}\right)
           \sqrt{\vec p_{\rm cm}\cdot\vec p_{\rm cm}\cdot+M^2}\, .\nonumber
\end{eqnarray}
A direct computation shows that these relative and center of mass variables commute and the coordinates within each class obey (\ref{poscomm}) and have the desired limit for large $M$.  From the start we would formulate a two body problem as
\begin{equation}\label{2body}
H=\frac{p^{\, (1)2}}{2m^{\, (1)}}+\frac{p^{\, (2)2}}{2m^{\, (2)}}+V(X^{\rm rel})\, .
\end{equation}
The use of these relative coordinated may be extended to many body situations. A many body Hamiltonian with interactions depending on the $\vec X^{\rm rel}$ is invariant under translations (\ref{translations}) generated by the total momentum.

\section{Minkowski Noncommuting Coordinates}\label{Minkowski}
In the previous section, by adding one {\em timelike} coordinate to three dimensional space we were able to identify $SO(3)$ invariant noncommuting coordinates with boost operators of the symmetry group $SO(1,3)$ of this extended geometry. Presently we will apply a similar procedure to $SO(1,3)$  invariant noncommuting coordinates, namely noncommuting space and time. By adding an extra time coordinate to $3+1$ dimensional Minkowski space, the noncommuting space-time operators will be represented by boosts of  the anti-de-Sitter group $SO(2,3)$, the symmetry group of the extended $2+3$ dimensional geometry. Again, as in Sect.~\ref{Euclidean} the discussion will be for $SO(1,3)$ coordinates embedded in $SO(2,3)$; an extension to $SO(1,d-1)$ embedded in $SO(2, d-1)$ is straightforward.

\subsection{A Representation of Poincar{\' e}-anti-de-Sitter Group}
We consider a representation of the Poincar{\' e}-anti-de-Sitter group acting on ``5-momenta" $p_T,p_\mu$, $\mu=0,\cdots,3$, preserving the metric $p^2=p_T^2+p_\mu p^\mu=p_T^2+p_0^2-\vec p\cdot\vec p$; $p_T$ and $p_0$ are the two time like momenta. For $p^2=M^2>0$ the Hilbert space consists of states $|p_0, \vec p\rangle$ with $p_T=\sqrt{\vec p\cdot\vec p+M^2-p_0^2}$. The norm of these states is taken as 
\begin{equation}\label{adsnorm}
\langle p_0',{\vec p}\, '|p_0,\vec p\, \rangle=\delta(\beta-\beta')\delta(\vec p-{\vec p}\, ')\, ,
\end{equation}
where $\beta$ is the angle between $p_0$ and $p_T$ or more specifically $p_0=\sqrt{\vec p\cdot\vec p+M^2}\sin\beta$ and $p_T=\sqrt{\vec p\cdot\vec p+M^2}\cos\beta$. Note that $|p_0|\le\sqrt{\vec p\cdot\vec p+M^2}$ or equivalently $p_\mu p^\mu\le M^2$. This restriction will be responsible for the eigenvalues of $X_0$, the time operator taking on discrete values.

The group algebra consist of Lorentz transformations ${\cal M}_{\mu\nu}$ and boosts ${\cal K}_\mu$ (${\cal K}_0$ is really an $O(2)$ rotation in the $p_T-p_0$ space). The commutation relations of the ${\cal K}$'s are
\begin{equation}\label{boostcomm2}
\left[{\cal K}_\mu,{\cal K}_\nu\right]=-i{\cal M}_{\mu\nu}\, .
\end{equation}
Analogous to the choice for the ${\cal K}_i$'s made in Sect.~\ref{Euclidean} for the representation of interest, in the present situation  an expression for the boosts that we shall use is
\begin{equation}\label{boost2}
{\cal K}_\mu= \sqrt{p_T}\left(-i\frac{\partial}{\partial p^\mu}\right)\sqrt{p_T}\, .
\end{equation}

\subsection{Coordinates}\label{coordianates2}
The usual commuting space-time coordinates can be related to their conjugate momenta by
\begin{equation}\label{ordcoord2}
x_\mu=-i\frac{\partial}{\partial p^\mu}\, .
\end{equation}
Noncommuting space-time coordinates are obtained by replacing (\ref{ordcoord2}) with
\begin{equation}\label{noncommcoord2}
X_\mu=\frac{1}{M}{\cal K}_\mu=\frac{1}{M}\left(M^2+\vec p\cdot\vec p-p_0^2\right)^{\frac{1}{4}} \left(-i\frac{\partial}
    {\partial p^\mu}\right)\left(M^2+\vec p\cdot\vec p-p_0^2\right)^{\frac{1}{4}}\, ;
\end{equation}
as previously, we denote commuting coordinates by lower case letters and their noncommuting counterparts by upper case ones.
The space-time coordinates $X_\mu$ satisfy 
\begin{equation}\label{coordcomm2}
\left[X_\mu,X_\nu\right]=\frac{-i}{M^2}{\cal M}_{\mu\nu}
\end{equation}
and in the limit $M\rightarrow\infty$, $X_\mu\rightarrow x_\mu$. The momentum-coordinate commutation relations are 
\begin{equation}\label{coormomcomm2}
\left[p_\mu,X_\nu\right]=ig_{\mu\nu}\frac{p_T}{M}\, .
\end{equation}

\subsection{Functions, Differentiation and Integration}\label{functions2}
As in the Euclidean case we make the correspondence between functions of commuting space-time coordinates and the noncommuting ones via the Fourier transform
\begin{equation}\label{f(x)tof(X)2}
f(x)=\int d^4q  \tilde{f}(q)e^{iq^\mu x_\mu}\rightarrow f(X)=\int d\mu_4(q )\tilde{f}(q)e^{iq^\mu X_\mu}\, ,
\end{equation}
with the invariant measure depending on whether $q^2$ is less than or greater than $0$,
\begin{eqnarray}\label{invmeas2}
d\mu_4(q)&=&\left[\frac{\sinh(\sqrt{-q^2}/M)}{\sqrt{-q^2}/M}\right]^3\, ,\ \ q^2\le 0\, ,\nonumber\\
d\mu_4(q)&=&\left[\frac{\sin(q/M)}{q/M}\right]^3\, ,\ \ q^2\ge 0\, .
\end{eqnarray}
There is, however, a caveat to this correspondence. For $q^2\ge 0$, $\exp(iq\cdot X)$ is an $O(2)$ {\em rotation} of angle $q/M$ between $q^\mu p_\mu/q$ and $p_T$; therefore $q^2$ must satisfy $q^2\le (M\pi)^2$. This also is related to the the fact that the time variable takes on discrete values.

As previously, differentiation can be generated by using the momentum vectors
\begin{equation}\label{diffdef2}
\partial_\mu f(X)=-i\left[p_\mu, f(X)\right]\, .
\end{equation}
In analogy with (\ref{noncommint}) integration of functions of noncommuting variables over all Minkowski space is defined as 
\begin{eqnarray}\label{integration2}
\int d^4X f_1(X)\cdots f_n(X)=(2\pi)^4& \int & d\mu_4(q_1)\cdots d\mu_4(q_n) \tilde{f}_1(q_1)\cdots \tilde{f}_n(q_n)
\times \nonumber\\
  &{}&\langle 0,\vec 0|e^{iq_1^\mu X_\mu}\cdots e^{iq_n^\nu X_\nu}|0,\vec 0\rangle\, ,
\end{eqnarray}
where the matrix element is taken in the state $|p_0=0,\vec p=\vec 0\rangle$. Some specific values of such matrix elements are:
\begin{equation}\label{eiqX2}
\langle 0,\vec 0|e^{iq^\mu X_\mu}|0,\vec 0\rangle=\delta(q_\mu)\, ,
\end{equation} 
as in (\ref{eiqX}), while for a product of two exponentials the result is
\begin{equation}\label{e1qq'X2}
\langle 0,\vec 0|e^{iq^\mu X_\mu}e^{-iq'^\mu X_\mu}|0,\vec 0\rangle = \delta(q_\mu-q'_\mu)\left\{
\begin{array}{l}
\left[\frac{{\sqrt{-q^2}}/{M}}{\sinh \left({\sqrt{-q^2}}/{M}\right)}\right]^3\, ;\ \  q^2\le 0 \, ,\\
{}\\
\left[\frac{{q}/{M}}{\sin({q}/{M})}\right]^3\,  ;\ \  q^2\ge 0\, .
\end{array}\right.
\end{equation}
For time-like $q$ the observation below (\ref{f(x)tof(X)2}) implies $-M\pi\le q\le M\pi$ and as in (\ref{eiqq'X}) the terms multiplying the $\delta$ functions cancel the corresponding ones in the invariant measure (\ref{invmeas2}).

\subsection{Position and Time Eigenvectors and Eigenvalues}\label{PosTimeEigen}
A complete, commuting set operators consists of $\hat r^\mu X_\mu$, with $|\hat r^2|=1$, and momenta orthogonal to $\hat r^\mu$. The study of these states has to be done separately for $\hat r^\mu$ timelike or spacelike and in latter case whether $(p+\hat r\hat r\cdot p)^2$ is larger or smaller than $M^2$. Lorentz transformations of $\hat r$ and $p$ do not mix these conditions. As there are varying subtleties in the procedure of obtaining the eigenvalues and eigenfunctions of  $\hat r\cdot X$ in the three cases we shall present each one in some detail. Special, simple examples of each of these situations are $\hat r=\hat z$ and $(p_x^2+p_y^2+M^2-p_0^2)\ge 0$, $\hat r=\hat z$ and $(p_x^2+p_y^2+M^2-p_0^2)\le 0$ or $\hat r$ along the time direction.

\subsubsection{$\hat r^\mu$ spacelike and $-(p+\hat r\hat r\cdot p)^2+ M^2\ge 0$ }
For $\hat r$ spacelike ($\hat r\cdot\hat r\le 0$), $p_T$ can be expressed in terms of $\hat r\cdot p$ and the magnitude of a vector orthogonal to $\hat r$,
\begin{equation}\label{r_spacelike}
p_T=\sqrt{-(p+\hat r\hat r\cdot p)^2+(\hat r\cdot p)^2+M^2}\, .
\end{equation}
We first consider the case $-(p+\hat r\hat r\cdot p)^2+M^2\ge 0$. 	It is useful to re-express $\hat r\cdot X$ in terms of a new variable $\eta$ defined by $\hat r\cdot p=\sqrt{-(p+\hat r\hat r\cdot p)^2+M^2}\sinh\eta$ and $p_T=\sqrt{-(p+\hat r\hat r\cdot p)^2+M^2}\cosh\eta$:
\begin{equation}\label{rdotX1}
\hat r\cdot X=%\frac{i}{M\sqrt{\cosh\eta}}\frac{\partial}{\partial\eta}\sqrt{\cosh\eta}=
\frac{i}{M}\left(\frac{\partial}{\partial\eta}+\frac{1}{2}\tanh\eta \right)   \, ,
\end{equation}
with $-\infty\le\eta\le\infty$. The $\delta$ function normalized solutions with eigenvalue $r$ are 
\begin{equation}\label{X1soln1}
\psi_{\hat rr}(\eta)=\sqrt{\frac{M}{2\pi\cosh\eta}}e^{-iMr\eta}
\end{equation}
(note: $d(\hat r\cdot p)=\cosh\eta\, d\eta$); this translates into
\begin{equation}\label{X1soln2}
\psi_{\hat rr}(\hat r\cdot p)=\sqrt{\frac{1}{2\pi p_T}}
 \left(\frac{p_T-\hat r\cdot p}{p_T+\hat r\cdot p}\right)^{\frac{iMr}{2}}\, .
\end{equation}

\subsubsection{$\hat r^\mu$ spacelike and $-(p+\hat r\hat r\cdot p)^2+ M^2\le 0$ }
In this case we find that $p_T^2\le (\hat r\cdot p)^2$ and we introduce the variable $\eta$ by $p_T=\sqrt{(p+\hat r\hat r\cdot p)^2-M^2}\sinh\eta$ and have to consider two possibilities for $\hat r\cdot p$, namely $\hat r\cdot p=\pm\sqrt{(p+\hat r\hat r\cdot p)^2-M^2}\cosh\eta$. The solutions are
\begin{equation}\label{X2soln2}
\psi_{\hat rr}(\hat r\cdot p)=\sqrt{\frac{M}{2\pi p_T}}
 \left(\frac{\hat r\cdot p-p_T}{\hat r\cdot p+p_T}\right)^{\frac{iMr}{2}}\, ,
\end{equation}
with $\hat r\cdot p$ in the two ranges $-\infty\le\hat r\cdot p\le -\sqrt{(p+\hat r\hat r\cdot p)^2-M^2}$ and $\sqrt{(p+\hat r\hat r\cdot p)^2-M^2}\le\hat r\cdot p\le\infty$.

\subsubsection{$\hat r^\mu$ timelike}\label{timeliker}
For $\hat r$ timelike $p_T^2=M^2-(p-\hat r\hat r\cdot p)^2-(\hat r\cdot p)^2$ with $(p-\hat r\hat r\cdot p)^2\le 0$. This time we define an angle $\eta$ by $\hat r\cdot p=\sqrt{M^2-(p-\hat r\hat r\cdot p)^2}\cos\eta$ and $\hat r\cdot p=\sqrt{M^2-(p-\hat r\hat r\cdot p)^2}\sin\eta$. For timelike $\hat r$, $\hat r\cdot X$ is a generator of rotations in the $\hat r\cdot p-p_T$ plane.
\begin{equation}\label{rdotX2}
\hat r\cdot X=\frac{i}{M}\left (\frac{\partial}{\partial\eta}-\frac{1}{2}\tan\eta\right)\, ,
\end{equation}
with solutions corresponding to eigenvalues $n/M$, where $n$ is any integer.
\begin{equation}\label{X3soln1}
\psi_n(\eta)=\frac{1}{\sqrt{2\pi\cos\eta}}e^{-in\eta}\, ,
\end{equation}
Written as a function of the momenta this wave function is 
\begin{equation}\label{X3soln2}
\psi_n(\hat r\cdot p)= \frac{1}{\sqrt{2\pi |p_T|}}\left(\frac{p_T-i\hat r\cdot p}{p_T+i\hat r\cdot p}\right)^{\frac{n}{2}}\, ;
\end{equation}
the range is $-\sqrt{M^2-(p-\hat r\hat r\cdot p)^2}\le\hat r\cdot p\le\sqrt{M^2-(p-\hat r\hat r\cdot p)^2}$ with $p_T$ both positive and negative. {\em The eigenvalues of coordinates in timelike directions are discrete and spaced by $1/M$}\/. In a state $\int d^4pf(\vec p)\psi_n(p_0)|p_0,\vec p\rangle$ (corresponding to $\hat r\sim t$) with $\int d\vec p|f(\vec p)|^2=1$, the expectation of the spatial position operator 
$X_i$ is linear in $n$, namely $\langle X_i\rangle=v_in/M$ with $v_i=\int d\vec p p_i|f(\vec p)|^2$. A discrete time has been obtained in formulations of space-time noncommutativity \cite{Chaichian:2000ia, Balachandran:2004yh}, where such noncommutativity involves more than the simple noncommutative plane.

\section{Free Scalar Field Theory}\label{FieldTheory}
Following the procedures of Sect.~(\ref{functions2}), we define a field by
\begin{equation}
\Phi(X_\mu)=\int d^4q \phi(q_\mu)e^{-iq\cdot X} \, ,
\end{equation}
making it an operator in both the Hilbert space on which $\phi(q_\mu)$ operates and in the $|p_0,\vec p\rangle$ Hilbert space. The latter makes it difficult to vary any proposed actions with respect to $\Phi(X_\mu)$, but it is possible to do a variation of the Fourier component $\phi(q_\mu)$. With differentiation defined in (\ref{diffdef2}) and integration in (\ref{integration2}) we propose the following action
\begin{eqnarray}\label{action}
A&=&\langle 0,\vec 0|\left\{-[p_\mu,\Phi^\dag(X)][p^\mu,\Phi(X)]-m^2\Phi^\dag(X)\Phi(X)\right\}|0,\vec 0\rangle\nonumber\\
&{}&\\
  &=&\int d^4q'd^4q \phi^\dag(q')\phi(q)\langle 0,\vec0|\left\{\left[p_\mu,e^{iq'\cdot X}\right]\Big[p^\mu,e^{-iq\cdot X}\Big]
      -m^2e^{iq'\cdot X}e^{-iq\cdot X}\right\}|0,\vec 0\rangle\, .\nonumber
\end{eqnarray}
For $q_\mu$ timelike (as expected, there are no solutions for $q_\mu$ spacelike) $\exp(-iq\cdot X)|0,\vec 0\rangle=
|Mq_0\sin(q/M)/q,M\vec q\sin(q/M)/q\rangle$ resulting in
\begin{equation}
\langle 0,\vec0|[p_\mu,e^{iq'\cdot X}][p^\mu,e^{-iq\cdot X}]|0,\vec 0\rangle=
    M^2\sin^2\frac{q}{M}\langle 0,\vec0|e^{iq'\cdot X}e^{-iq\cdot X}|0,\vec 0\rangle\, ;
\end{equation}
the matrix element in the above can be found in (\ref{e1qq'X2}).
Setting the variation of (\ref{action}) with respect to $\phi(q)$ to zero yields the dispersion relation
\begin{equation}\label{dispersion1}
q^2=M^2\arcsin^2\frac{m}{M}\, .
\end{equation}
As $0\le q^2\le \pi^2M^2$, we obtain, for each $m^2$, two solutions $q^2=m^2_l$ and $q^2=m_l^2$, with $m_h>m_l$. For $m/M$ small we find in addition to the usual solution $m_l=m$, the solution $m_h=\pi M-m$; the mechanism responsible for having two solutions is reminiscent of fermion doubling \cite{fermdoubl} in lattice field theories, except that in the present situation the mass of the additional state is comparable to $M$ and the energy is always large; as $M$ goes to infinity this extra mass decouples. Importantly, we note that solutions of the field equations exist {\em only for $m\le M$}. 

Aside from the doubling question, we may relate $\phi(q_\mu)$ for $q_0=+\sqrt{\vec q^2+m_{l,h}^2}$ to annihilation operators and for $q_0=-\sqrt{\vec q^2+m_{l,h}^2}$ to creation ones and express the field operators as in commuting field theories with $x_\mu$ replaced by $X_\mu$.

\section{Remarks and Conclusions}\label{Conclusions}
Identifying coordinates as boost operators of the symmetry group of a space with one timelike coordinate added to the space of interest allows us to replace usual commuting coordinates or coordinates and time can by noncommuting ones. These dynamics are rotationally, respectively, Lorentz invariant. As discussed in the introduction, the latter claim could be valid for noninteracting theories as usual formulations of interactions using time ordered products require that commutators of operators vanish for space-like separations. Many body interactions can be  formulated in a way to make them invariant under rotations {\em and} translations. Upon the introduction of  time resulting field theories are invariant under the full Poincar\'e group. The noncommutativity is governed by a parameter $M$ and in the large $M$ limit we recover ordinary quantum mechanics or field theory. Noteworthy results are:
\begin{itemize}
\item
The Heisenberg commutation relation between position and momentum is modified for distances smaller than $1/M$ and/or momenta larger than $M$.
\item
Eigenvalues of the coordinate operator in the time or any timelike direction are discrete and spaced by $1/M$.
\item
Wave packets cannot be localized to $\sqrt{\langle \vec X^2\rangle}$ less than $1/M$.
\item
A free field theory with mass $m$ has a solution only for $m$ less than $M$. 
\end{itemize}
The last two results prevent us from packing a mass greater than $M$ into a space of radius less than $1/M$; if the noncommutativity parameter $M$ is smaller than $M_{\rm Planck}$, this would preclude the existence of pointlike black holes, that is black holes with radius $\sim 1/M_{\rm Planck}$ and mass $\sim M_{\rm Planck}$. 

Two topics that have not been addressed are the UV-IR connection \cite{Minwalla:1999px} and the violation of causality, when time is included. Such violations occur for noncommuting space-time \cite{Seiberg:2000gc} or, even when time is a continuous variable commuting with the space coordinates \cite{Greenberg:2005jq}.

\end{document}